# A 20kA ±0.12% Pulsed Power Supply for the PIP-II ORBUMP Injection Dipole Magnets

H. Pfeffer, M. Davidson, D. West, S. Garcia-Schiefelbein, G. Lolov, R. Rivera-Colon
Fermi National Accelerator Laboratory
Batavia, USA
pfeffer@fnal.gov, mattd@fnal.gov, dwest1@fnal.gov, sschiefe@fnal.gov, glolov@fnal.gov, ramfis@fnal.gov

***Abstract*— Fermilab has embarked on its next evolution for future neutrino experiments through the Proton Improvement Plan II (PIP-II). The injection system now requires an increase from a 15 Hz to a 20 Hz injection rate with approximately a 30% increase in the field strength from the original Orbital Bump (ORBUMP) dipole magnets in the Booster ring [1]. The four series connected dipoles will require 3 kV, 20 kA, 660 µs pulses at the aforementioned injection rate with a flatness of ±0.12%. This is achieved through a dual-capacitor pulsed power supply (PPS) comprised of ten parallel cells and must last for 25 years. The supply utilizes high-voltage film caps for the rise and fall portion of the waveform; however, it switches over to a 6 F electrolytic capacitor bank for the flattop portion to meet flatness. The supply utilizes energy recovery to reduce high voltage charging supplies and required power to operate.**

## I. Introduction

The PPS for the new ORBUMP magnets takes the form of a standard unipolar capacitor discharge pulser but utilizes three capacitor banks: the pulse and recovery banks for the rise and fall portions of the waveform and a flattop bank for 660 µs flattop portion. The PPS utilizes separate IGBTs for the pulse and recovery operations while the flattop capacitor bank is switched naturally through a diode. The PPS is further divided into 10 cells to allow construction of each cell with single semiconductor devices (i.e., IGBTs and various diodes) that support operation in the 2 $kA_p$ and 3 $kV_p$ region and to segment the energy storage into smaller isolated sections. These choices allow the PPS to meet key specifications of operating at 20 Hz, with a peak voltage of 3 kV, peak current of 20 $kA_p$ maximum flattop length of 660 µs, a flatness of ±0.12%, maximum rise time of 450 µs, maximum fall time of 250 µs, and maximum rms current of 3 kA.

Four series connected dipole magnets, located in the Long 11 straight section of the Fermilab Booster, serve as the load presenting an impedance of 24.8 µH and 1.3 mΩ. An 80 ft, 5-layer stripline and 50 ft of 50, RG-220 cables comprise the distribution from the power supply to the load. This contributes an additional 500 nH, mostly from the stripline, and an additional 820 µΩ. The stripline allows compact power delivery within the highly occupied Booster enclosure, while the cables allow flexibility up through the service building penetrations.

The present day Linac's ORBUMP power supply is a pulse forming network (PFN) that operates at 15 Hz, 17.5 $kA_p$ for a duration of 50 µs with a flatness of ±0.5%. A PFN was considered initially, however, this was found to be unfeasible due to the difficulty in achieving the stringent flattop specification without many stages and tuning elements and poor load matching requiring at least 6 kV.

## II. Power Supply Theory and Design

### A. Theory of Operation

The operation of the PPS will be described through a single cell as seen in Fig. 1, which is equivalent when considering the entire supply. The pulse capacitor bank (C1) will be initially charged to 2.8 kV and the flattop capacitor bank (C3) to 60 V for 20 kA operation. A fire command issued causes the main pulse switch (Q1) to close, which connects the load to C1 through the fault choke (L2) and provides the transition to peak current under 450 µs (defined as time from fire to beam injection). As C1 discharges to the voltage level of C3, diode D1 permits the transition to the C3 bank. This bank supplies current for the duration of the flattop until Q1 turns off, up to 660 µs. Once Q1 is open, energy stored in the load is transferred to the recovery bank (C2) via diode D2. This allows the current to decay under 300 µs (90%-10%) resulting in a C2 voltage of -2.5 kV due to the recovered charge. To recover the energy from C2 back into C1, the same pulse topology is used: Q2 transfers energy from C2 to L1 and once peak current is reached, Q2 opens allowing energy to transfer from L1 to C1 through diode D3. Salient waveforms are shown in Fig. 2 for the pulse and recovery stages. Top-off charging for both C1 and C3 complete the 50 ms period.

### B. Charging and Crowbar

All capacitors for the pulse (C1) and flattop (C3) banks are charged in parallel by dedicated supplies for each bank. The pulse capacitors are charged through parallel commercial capacitor charges (6-10kj/s range), and the flattop capacitors are charged through parallel commercial DC supplies (10-15kW range).

The choice of paralleling ensures common power ranges, avoiding lock-in to a specific type, and potential high uptime with N+1 designs and easier replacement. Utilizing charge recovery for C1 reduces charging power by 80%.

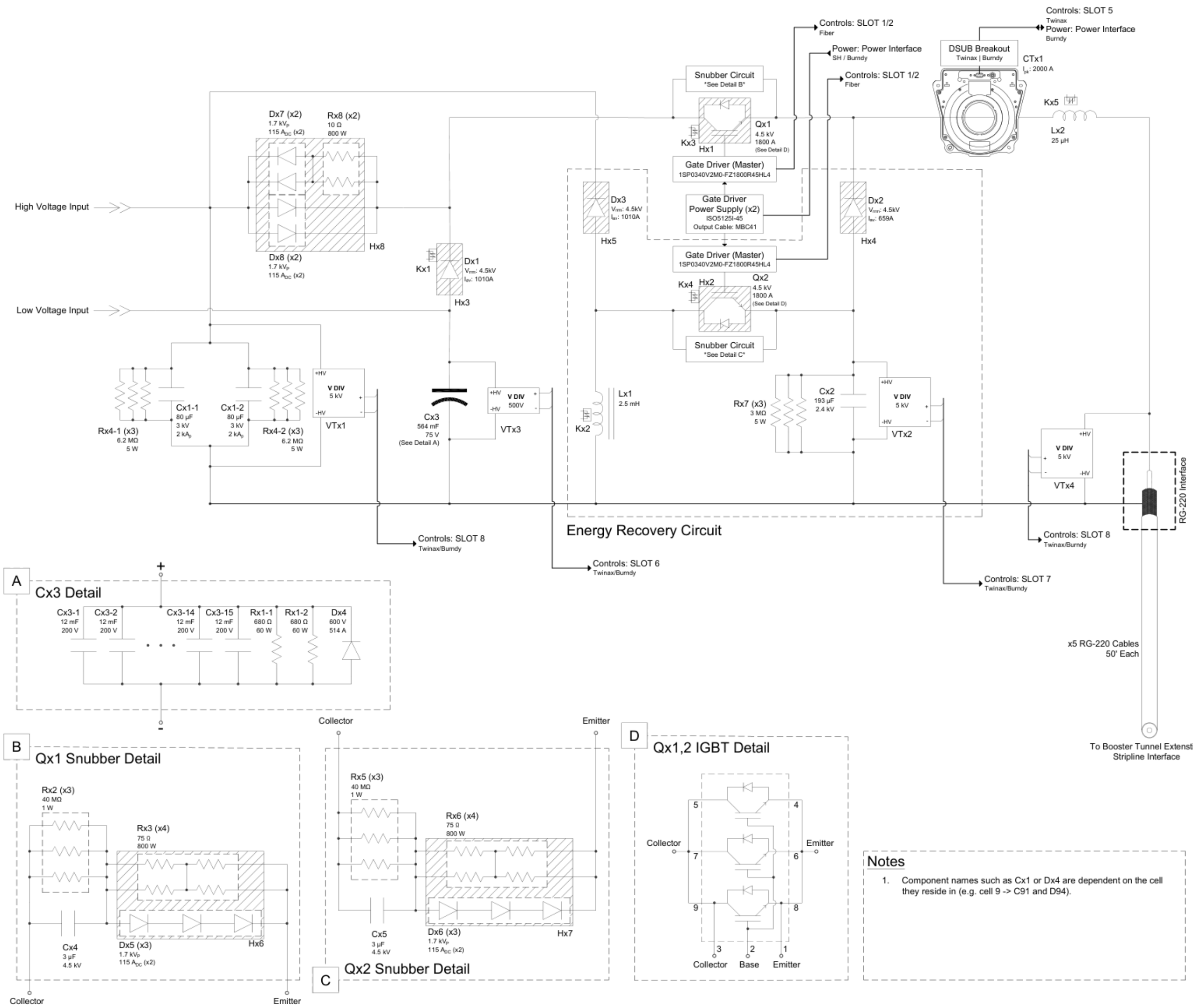

*Fig. 1. Schematic showing elements for a single cell.*

The charging is managed through controls, which defines the repeatability thus reducing the performance requirements of the commercial supplies. The charging supplies are each diode isolated before combining to the respective charging bus. Each cell is also diode isolated as this localizes failures and prevents the full bank discharging into potentially shorted adjacent banks or charging supplies.

Each cell capacitor bank (pulse, flattop, and recovery) is resistively tied to their respective discharge bus through disc style resistors. Each bus is tied to redundant Ross relays at each end of the PPS to discharge the high voltage banks in 1-2 seconds, longer for the flattop bank. The redundancy provides protection not only in the failure of one of the relays, but if the bus becomes open at one point, the relays can still discharge all capacitor banks.

## C. Semiconductors

As previously mentioned, IGBTs are used as the controllable switching elements for the pulse and recovery stages, which provide variable flattop length and the ability to turn off early in the case of a fault. The choice of the IGBT was the main determining factor for 10 cells. The Infineon FZ1800R45HL4, driven through a dedicated commercial gate driver, allows at least 1.5x ratings for peak operating voltage and current. Additionally, the dc stability voltage should be heeded as some cases may incur sustained voltage blocking. The main thermal limitation is to prevent cyclic temperature swings from exceeding 5 °C, which is accomplished with 2 gpm water cooling and a low thermal impedance from the IHM-B package. This design restriction conservatively aligns with projected lifetime data for the IGBT 4 devices: 1E+10 pulses possible with ΔT of 10 K [2]. An RCD snubber is utilized for the turn-off transient, where low inductance resistors and diodes take advantage of the chill block used for the IGBT; the capacitor is a custom device from ECI to ensure comfortable margin with operating voltage, peak current, and dv/dt. The resistance is higher than typical at 70 Ω, which aids in reducing the load current undershoot during the reverse recovery of D2.

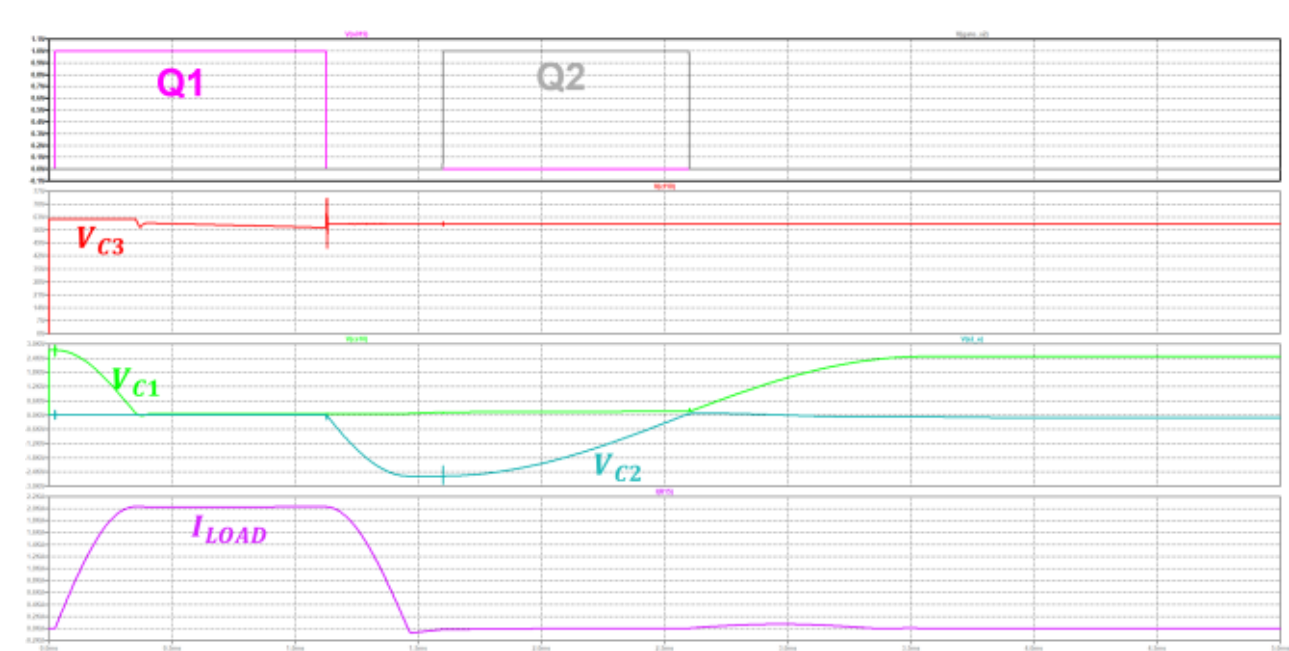

*Fig. 2. Simulation waveforms showing capacitor voltages and load currents with respect to switch timing.*

Diodes are chosen as puck devices and are rated with similar margins. These devices are also water cooled to ensure high reliability; however, this requires the use of low conductivity water as the chill plates are at potential. Diodes D1 and D3 utilize the IXYS M1010NC450, which are also required to meet the same dc stability voltage requirement as the IGBT. Diode D2 utilizes the IXYS M0659LC450, where more attention is paid to reverse recovery current as the PPS is limited to -60A per cell as required by the specification to prevent displacing the beam orbit.

### D. Magnetics

The energy recovery choke enables recovery from C2 to C1 and was designed to be 10x the load inductance to reduce peak currents in the recovery phase. This presents as a 2.5mH choke (per cell) designed at 720 $A_p$, 100 $A_{rms}$, and 3kV. Losses are targeted to be below 700 W with a max operating temperature of 130 °C; techniques to limit core loss are encouraged such as distributed gaps.

The fault choke provides a mechanism for detecting load short circuit faults within an allowable time frame before normal peak currents are reached through Q1. The value is chosen to be 10x less than the load inductance to limit the impact on the C1 operating voltage. This is designed as a 25 µH choke (per cell), 2 $kA_p$, 300 $A_{rms}$, and 3 $kV_p$ choke where operation up to 6 $kA_p$ 100 times is permitted for faults. Losses are limited to 350 W, which makes core-based designs difficult, so air-core designs are acceptable provided the stray field is kept below 100 Gauss near cabinet walls.

Magnetics in a pulsed environment are difficult due to the magnetic forces from the peak cyclic current. Achieving a 25-year lifetime with 20 Hz pulse rate means the magnetics must endure >10 gigacycles; mechanical construction should be robust and potting is highly encouraged.

### E. Capacitors

Capacitor banks C1 and C2 are designed to provide a reasonable rise and fall time, which balances operating voltage with the rms current through the magnet. The pulse bank was designed to be 160 µF per cell, requiring a 3 $kV_p$, 2 $kA_p$, 130 $A_{rms}$ bank. Each bank is comprised of two parallel capacitors, primarily dictated by limiting the current rating per cap to below 70 $A_{rms}$, which makes realizing single bushing film capacitors practical for this application.

To maintain the flattop current, a 360-600 mF bank per cell operating around 65 V is utilized. The capacitance can vary due to the non-constant load impedance, where a smaller bank can counteract the variation. Electrolytics are an attractive choice due to the size and cost penalties of a film implementation; however, electrolytics in long-life pulse applications are a cause for concern. These must be specified for high ripple currents, low ESR, and high temperature rating. The bank is comprised of at least 15 capacitors, limiting peak and rms currents and lowering effective ESR. It is determined as the bank approached 25-30 capacitors, the probability of failure in a single device becomes the limiting factor alongside diminishing returns. Temperature is exponentially correlated with lifetime based on the Arrhenius law [3], where a reduction of 10 °C can result in a doubling of life [3,4], therefore capacitor construction utilizing the case for cooling (e.g., stud mount, extended foil construction) is preferred; forced convection as needed. Operating in partial discharge (<15V per pulse) and with a voltage derating of 60-80% also contributes to longevity. To ensure reliability over the lifetime of the PPS, the electrolytics are placed on a 10-year program, with spares replenished every 5 years to limit storage degradation; reforming is performed before replacement.

### F. Capacitor Damping

As the energy source transfers from C1 to C3, the ESL of each bank forms a resonant circuit, causing large circulating currents to develop. This will cause increased heating in C3 and is observable on the load adversely affecting flattop regulation. An added damping circuit will squelch the oscillation while minimizing overall resistive losses. This is realized by a lossy element for C3 to C1 currents, a bypass diode, and blocking diode as seen in Fig. 1. The combination of the diodes allow C1 to discharge as normal after the initial transient; C3's discharge and C1's recharge are unaffected by this circuit. The bypass diode is required to be either a fast recovery or Schottky device; the diode voltage requirements are small but dependent on resistor choice and must withstand the initial current rise.

## III. Controls, Measurement, and Fault Considerations

### A. Controls Design

The controls design follows the platform of the LBNF modulator [5] due to the multi-cell architecture and imbalance detection circuits but is adapted to the smaller cell count and differing transducers used in the ORBUMP PPS. The state logic utilizes four different logic flows: normal, regulation fault, major fault, and severe fault. Severe faults impacting human safety or presents imminent destruction activates crowbars and opens the charging supply contactor; major faults stop operation but only inhibit charging to avoid long recharge cycles; regulation faults pull the beam permit but otherwise continue pulsing to maintain temperature; normal maintains the beam permit.

The controls platform will interface with EPICS (Experimental Physics and Industrial Control System), which will provide readback of salient values and provide parameter setpoints, namely: the current value and flattop width; an external trigger provides a FIRE command. The direct control variables for the PPS are the final charge setpoints for each capacitor bank (pulse (C1) and flattop (C3)). The C1 voltage largely determines the intended operating current, and is mapped to the current setpoint, while C3 tracks to C1's voltage to maintain the flatness of the pulse. A feedback loop is applied to C1's voltage to maintain the pulse-to-pulse current stability, which mainly compensates for thermal variations. Bank C3 has no direct feedback, although the optimal operating voltage for a given current setpoint could be investigated using the flattop current $\Delta I$ as a cost function to minimize.

Two sets of 10 IGBTs are the controllable elements of the PPS: 10 IGBTs providing the main pulse current and 10 IGBTs allowing energy recovery. Certain fault conditions will also remove power from the Ross relays and charging power supply contactor. These are powered through an independent series relay chain for safety reasons; however, the control system serves as a series element alongside the PPS and charging door switches, an Electrical Safety System (ESS) permit (for tunnel safety), and E-STOP buttons.

### *B. Measurement and Detection*

Compensated resistive voltage dividers comprise the voltage monitors for C1 voltage, C2 voltage, C3 voltage, and load voltage. An internal design is used to achieve a bandwidth up to 1 MHz and compatibility with long cable lengths; commercial dividers are also a possibility. A LEM DCCT is chosen for the output current measurement as CTs will have challenges related to saturation and tilt exceeding the flattop regulation value (although this may be compensated for). The output of the DCCT is a current source to increase immunity against noise and cable resistance.

Vortex flow transducers are used in each cell's water path to ensure localized problems (such as blockages or trapped air) are caught before causing thermal issues. A differential pressure switch monitors the main water supply and serves as a redundant trip to the localized flow transducers.

Remaining analog sensors are for ambient temperature and humidity (for both the PPS and charging rack) and current transducers for total charging current and ground currents. Contact-based sensors (switches) comprise the remaining detection elements for alerting the controls of over-temperature conditions (e.g., bus work, semiconductor elements), capacitor over-pressure, ambient over-temperature (redundant), and smoke detection. Similar contact-based over-temperature detection is used for the load and stripline.

Auxiliary faults are also monitored through simple status (contact or fiber) such as with charging supply failure (detected from an in-house Multiple Power Supply Interface Controller (MPSIC)), controls power supply failure, and power loss failure (detected through the controls UPS).

### *C. Fault Considerations for a Single Cell*

Each cell must protect itself from a load short circuit and the prevention and detection are accomplished with each cell's fault choke. Expanding on Section II-D, in the worst-case scenario (short circuit at output before cables at turn on), the fault choke limits the di/dt to below 115 A/μs, which allows a response within 18 μs before peak operating current is reached; an extended response time is permitted up to 6 $kA_p$ with the peak fault choke current. The initial detection occurs by subtracting the output voltage from C1's voltage to check if the voltage difference across L2 exceeds a threshold. The DCCT would also detect an over current as a redundant catch and digitization of the current measurement could allow for software di/dt detection as well.

Switch failures are primarily detected through the IGBT gate drivers, which provides an active signal for device powered/ready and an acknowledge signal for each trigger. Notably, early turn-off of Q1 during the initial current rise will cause up to a 30% increase in $V_{CE}$ due to the remaining charge on C1 and recovered charge on C2. High operational margins for diodes ensure very low failure rate, but should that occur, faults are mostly isolated to the cell.

Individual component failures are primarily detected through the imbalance and window circuits described in the next section. However, being that C3 is an electrolytic bank, a bypass diode is in parallel with the bank to prevent reverse voltage exceeding one diode drop. This will occur if the C3 bank does not have sufficient charge to support the flattop. Threshold detection is used to ensure capacitor states are correct before proceeding such as C3 charged and C2 discharged before firing Q1 and C1 discharged before firing Q2, which prevents reversals and over voltages.

### *D. Fault Consideration for 10 Cells*

Each voltage, current, and water sensor are fed to separate imbalance boards, which compare an individual cell's measurement with the instantaneous average of the PPS. Absolute upper and lower limits are also monitored against the average signal. This captures a majority of faults related to open or shorted conditions, switch no fire, parallel cable faults, or large deviations in component values (e.g., partial dielectric failure in capacitors, reduced inductance through turn to turn short). The sensitivity of the detection is dependent on the component matching between components in each cell.

The combined charging current is measured for each supply to avoid charging into a fault by implementing overtime trips. Faster detection can be implemented by digitizing and evaluating the change in the respective capacitor bank average voltage with respect to the measured current value.

Ground faults are captured through the single point ground within the PPS by either a CT or shunt measurement with threshold detection. Dedicated ground fault return path is provided alongside the RG-220 cables and stripline to force fault currents, and also capacitive currents, back to the PPS cabinet.

## IV. Prototype and Results

A single cell prototype was constructed to vet the topology and veracity of the simulations performed through LTSpice. The film caps limited the peak current to 1.7 $kA_p$ and the repetition rate to 3 Hz or 20 Hz bursts (one second on to five seconds off). The output is connected via five parallel 50 ft

RG-220 cables to a 250 µH load, which represents the load impedance for a single cell. Simple analog limit checks are monitored, such as C1 overvoltage and load overcurrent, and the regulation for the prototype is achieved through a set reference for each charging supply, which was able to maintain acceptable precision. While some drift was observed, this will be addressed with the final regulation system.

Operation of the prototype confirmed expected performance of the IGBT snubber circuit; rise and fall time performance, measured at 400 µs and 230 µs, respectively; load current undershoot, measured at -40A; and flattop performance, measured repeatably within ±1.5 A for a 1650 A pulse (see Fig. 3).

Investigation allowed refinement of Q2 switching times, which resulted in delaying Q2's turn-on until D3's current fully decays, otherwise a large transient occurs. Additionally, turning off Q2 early (leaving some voltage on C2), reduces a post-pulse current bump within acceptable limits. Results indicated that a custom snubber cap was required to maintain good margin with dv/dt limits, an RC termination network was required for the load due to reflections, and the need for the damper circuit discussed in Section II-F was confirmed.

The results of the single cell prototype confirm the topology can meet the technical requirements for the ORBUMP magnets. A two-cell full scale prototype is underway to examine the updated design at full current and rep rate, while allowing investigation into performance and faults with parallel cells.

## ACKNOWLEDGEMENT

Kevin Roon for initial electromechanical design and concepts, Austin Sahr and Alex Saracino for help with the prototype construction, Chris Jensen and Matt Kufer for constructive feedback during reviews, and Nick Gurley for assisting with controls design and controls reviews.

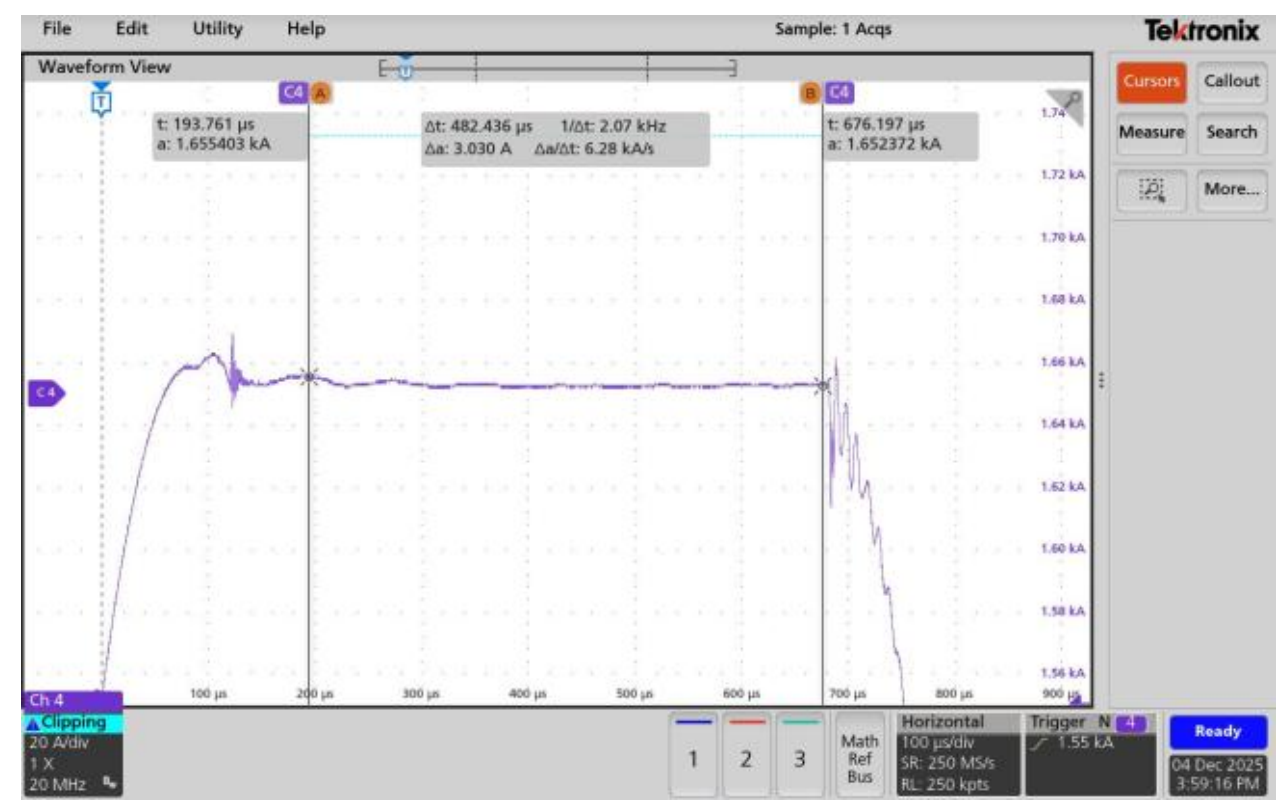

*Fig. 3. Oscilloscope capture showing load current flattop performance at 1.65 kA for 480 µs.*